\documentclass[12pt,preprint]{aastex}

\shorttitle{GJ 841B--A New CH White Dwarf}
\shortauthors{Vornanen et al.}

\begin{document}


\title{GJ 841B---The Second DQ White Dwarf with Polarized CH Molecular Bands}


\author{T. Vornanen\altaffilmark{1},
        S.~V. Berdyugina\altaffilmark{2},
        A.~V. Berdyugin\altaffilmark{1},
        V. Piirola\altaffilmark{1}
}



\altaffiltext{1}{Department of Physics and Astronomy, University of Turku, FIN-21500 Piikki\"{o}, Finland; emails: tommi.vornanen@utu.fi, andber@utu.fi, piirola@utu.fi}
\altaffiltext{2}{Kiepenheuer-Institut f\"{u}r Sonnenphysik, D-79104 Freiburg, Germany; email: sveta@kis.uni-freiburg.de}


\begin{abstract}
We report a discovery of the circularly polarized CH $A^2\Delta$--$X^2\Pi$ and $B^2\Sigma^-$--$X^2\Pi$ molecular bands in the spectrum of the DQ white dwarf GJ 841B.  This is only the second such object since the discovery of G99-37 in the 1970s. GJ 841B is also the first WD to unambiguously show polarization in the C$_2$ Swan bands. By modeling the intensity and circular polarization in the CH bands we determine the longitudinal magnetic field strength of 1.3$\pm$0.5\,MG and the temperature of 6100$\pm$200\,K  in the absorbing region. We also present new observations of G99-37 and obtain estimates of the magnetic field strength 7.3\,$\pm$0.3\,MG and temperature 6200$\pm$200\,K, in good agreement with previous results. 
\end{abstract}





\section{Introduction}

DQ white dwarfs (WDs) constitute a subclass of DB helium-rich WDs and are characterized by neutral and molecular carbon features in their spectra. At their temperatures between 4000\,K and 13,000\,K carbon dredged up from the core \citep{koe82} can form molecules, the most abundant of which are C$_2$ (without the presence of hydrogen), CH (in the presence of minute amounts of hydrogen in the atmosphere), and perhaps under certain conditions C$_2$H \citep{sch95}. The presence of the latter in the spectra of so-called peculiar DQ WDs has been however seriously questioned by \citet{hal08}. As with all WDs, a certain percentage of DQ WDs are magnetic. For isolated DA- and DB dwarfs this is about 5\%\ \citep[][before the Sloan Digital Sky Survey]{wic00}. However, it is not known how many of the DQ WDs are magnetic. Currently five such stars are known; LP790-29 \citep{lie78, wic79, bue99}, LHS2229 \citep{sch99}, SDSS J1113+0146 \citep{sch03}, SDSS J1333+0016 \citep{sch03}, and G99-37 \citep{ang74}. Of these three G99-37 shows polarization in the CH molecular bands, while the other four are members of a group of peculiar DQs with broad absorption bands which are strongly polarized. With the addition of our newly discovered sixth magnetic DQ WD, 4\% of 151 confirmed DQ WDs listed in the SIMBAD database are magnetic. One reason for this lack of objects with known magnetic fields is that spectra of these cool carbon WDs are complicated by molecular bands, which makes it challenging to deduce and measure a magnetic field.

\citet{ang74} pioneered a method to measure the magnetic field strength in \object{G99-37} from the circularly polarized CH $A^2\Delta$--$X^2\Pi$ molecular band, which they had discovered. This method was further developed by \citet{ber05} and applied to new sensitive spectropolarimetric measurements of G99-37 to obtain an improved estimate of its magnetic field \citep{ber07}. The new analysis has also discovered circular polarization in the CH $B^2\Sigma^-$--$X^2\Pi$ system. Thus, for over 35 years G99-37 was the only known WD with CH molecular bands and the remarkable polarization (up to about 10\%) associated with them. 

Here we report the discovery of a second DQ WD with circularly polarized CH spectral features, GJ 841B, and provide the first estimate of its magnetic field. We have also obtained new, high signal-to-noise ratio (S/N) measurements for G99-37 in 2005 and 2008. We compare them with those of 2003 published by \citet{ber07} in order to check for possible long-term variability.

\section{Observations}

Observations of \object{GJ 841B} and G99-37 were carried out in a spectropolarimetric mode with the now-retired FORS1 instrument \citep{app98} at the Very Large Telescope (VLT) on the nights of 2008 November 15 and 16, together with a sample of other cool carbon WDs (results obtained for them will be presented in a separate paper). The observations were done with the grism 600B+12 which covered a wavelength range 330--621\,nm with the spectral resolution of $R$=780. The stars were always in the center of the field of view.
In addition, G99-37 was observed on 2005 December 12, with the same instrumental setup. The spectra were reduced using standard IRAF routines.

The observed wavelength interval includes several molecular features from CH and C$_2$. To achieve the polarization sensitivity of $\pm$0.2\% (S/N$\sim$700), we concentrated our efforts on circular polarization measurements (Stokes $V$), where we also expected the largest signal based on the previous data for G99-37. Nevertheless, we also obtained a measurement in the linear polarization (Stokes $Q$ and $U$) for GJ 841B. In contrast to Stokes $V$, no significant signals could be seen at the level of $\le$0.1\%. A cross-talk from Stokes $V$ to linear polarization which is a known issue in FORS1 \citep{bag09} is also not seen. Thus, this provides us with the upper limit.

The spectra of G99-37 from the years 2005 and 2008 are very similar, so we focused our analysis on the later one. These spectra allowed us to calibrate the sign of polarization in GJ 841B. 

\section{Results}

Our analysis of the data is based on the theory of polarization in diatomic molecules elaborated by \citet{ber05}. Molecular bands show broad-band net polarization due to the Paschen--Back effect i.e., when an internal coupling of the molecular angular momenta is weaker than their interaction with an external magnetic field. Following the method developed by \citet{ber07} for interpretation of the G99-37 data, we calculated Stokes spectra from theoretical Zeeman patterns (transition strengths and wavelength shifts) for the two electronic systems of CH: $A^2\Delta$--$X^2\Pi$ and $B^2\Sigma^-$--$X^2\Pi$ as well as for blending C$_2$ bands. Our model employs an analytical solution of the polarized radiative transfer equations, the so-called Unno--Rachkovsky solution, based on the Milne--Eddington stellar atmosphere. The spectral synthesis involves over a million Zeeman transitions. Their characteristics are discussed in detail by \citet{ber07}. The only difference to that paper is that here we consider a homogeneous magnetic field along the line of sight (instead of a dipole), as we analyze Stokes $V$ only.

\subsection{GJ 841B}

\begin{figure}
\resizebox{14cm}{!}{\includegraphics{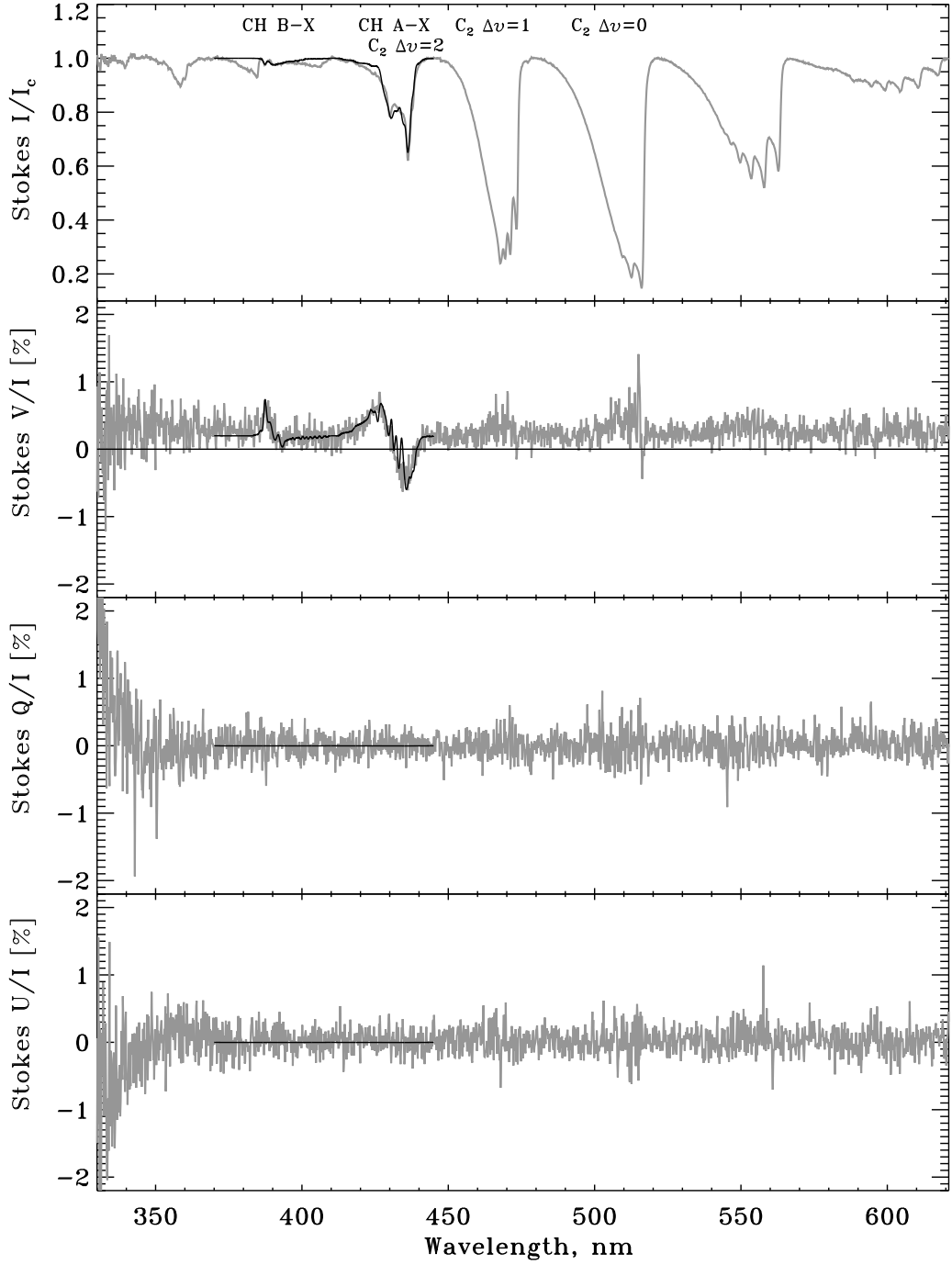}}
\centering
\caption{Intensity (top) and Stokes $V$ (bottom) of GJ 841B. Gray line is observation, and black line is the model with $T=6100$\,K and the longitudinal field $B=1.3$\,MG.} 
\label{fig:GJ841B}
\end{figure}

GJ 841B (BPM 27606, WD 2153-512) has remarkably strong absorption bands of C$_2$ with a central depth down to 20\% of the continuum level (see Figure~\ref{fig:GJ841B}). This object was known for many years as a non-magnetic DQ WD with the strongest C$_2$ lines in the spectrum \citep[see][]{wic79, sch99}. \citet{koe82} modeled C$_2$ bands and energy distribution in GJ 841B and estimated the temperature on its surface to be 7600$\pm$500\,K. Later, \citet {weg84} from their model calculations obtained a lower value of 7000\,K. It seems that so far no one considered the possibility that some features in its spectrum could be blends of C$_2$ and CH lines. 

Our new high S/N spectropolarimetry has revealed that the absorption feature at 430\,nm is the CH $A$--$X$ system blended with the C$_2$ Swan $\Delta v=2$ absorption bands centered at 438\,nm. In contrast to G99-37 where these C$_2$ bands are hidden in the broad CH absorption, here the CH bands being weaker are barely visible in the blue wing of C$_2$ (Figure~\ref{fig:GJ841B}). Such a situation may be typical for DQ dwarfs, and a more detailed analysis of their spectra is needed to reveal perhaps traces of CH. However, most remarkably CH is visible in Stokes $V$ where it cannot be missed due to the striking antisymmetric profile, as in G99-37, but with a lower amplitude, from $-$0.5\%\ to +0.7\%.
 
The above conclusion is strengthened by the presence of the CH $B$--$X$ system at 390\,nm which is also well hidden in Stokes $I$ between two C$_2$ bands of the $C^1\Pi-A^1\Pi$ Deslandres-d'Azambuja system at 405\,nm ($\Delta v=-1$) and at 385\,nm ($\Delta v=0$). Other bands of this system are also visible at 360\,nm ($\Delta v=1$) and 330\,nm ($\Delta v=2$). They are not expected to be polarized as explained by \citet{ber07}. The Stokes $V$ profile of the CH $B$--$X$ system peaks at about +0.6\%. Again its shape is reminiscent of that in G99-37, suggesting unambiguously its origin due to CH.

Another similarity to G99-37 is that the continuum is also polarized, but at a lower level, about 0.2\%. In addition we detect, for the first time in WDs, polarization associated with the C$_2$ Swan bands: Stokes $V$ peaks at about 0.5\%, and Stokes $Q$ perhaps at $\le$0.1\%. This gives us hope that some other DQ dwarfs with very strong molecular carbon bands may show similar features if observed with a high polarimetric sensitivity.

Using our model, we searched for the best-fit solution simultaneously for Stokes $I$ and $V$ of the two CH systems. The result is that the strength of a longitudinal magnetic field $B=1.3\pm0.5$\,MG and temperature $T=6100\pm200$\,K (see Figure~\ref{fig:GJ841B}). The uncertainties come from the fitting procedure; within these limits the fits are equally good. A weaker magnetic field as compared to G99-37 is in agreement with the lower CH polarization \citep[see the dependence on the magnetic field strength in][]{ber07}. 

We also included the contribution from the C$_2$ $\Delta v=2$ bands, which noticeably affect the intensity and polarization of CH in the red wing.
Some remaining deviations from the observations can be due to still unaccounted C$_2$ blends. Also, the distribution of the field on the surface of this WD may not be homogeneous as assumed, but it may be very close to that since the overall profile shapes are well reproduced. The assumption on longitudinal field seems to be acceptable, because there is no significant linear polarization in the CH bands. When assuming a dipole field configuration \citep[as in G99-37][]{ber07}, the noise level provides the limit on the dipole inclination with respect to the line of sight $\gamma<20^\circ$. With a larger inclination some signal would be visible in Stokes $Q$ and $U$. We believe that a more complex field distribution and, hence, possible polarization cancelations could be excluded, because we were able to fit simultaneously Stokes $I$ and $V$ without introducing any magnetic filling factor.

Further constraints on the field geometry can be provided by modeling polarization in the Swan bands of C$_2$. However, we experienced some difficulties to obtain a consistent, equally good fit for them. The problem of discrepancy between the calculated and observed relative depths of different Swan bands in some carbon DQ WDs was also mentioned by \citet{weg84} and more recently by \citet{duf05}. Thus, it appears that this problem needs a thorough investigation and we will approach it in a separate paper.

\subsection{G99-37}

\begin{figure}
\resizebox{12.0cm}{!}{\includegraphics{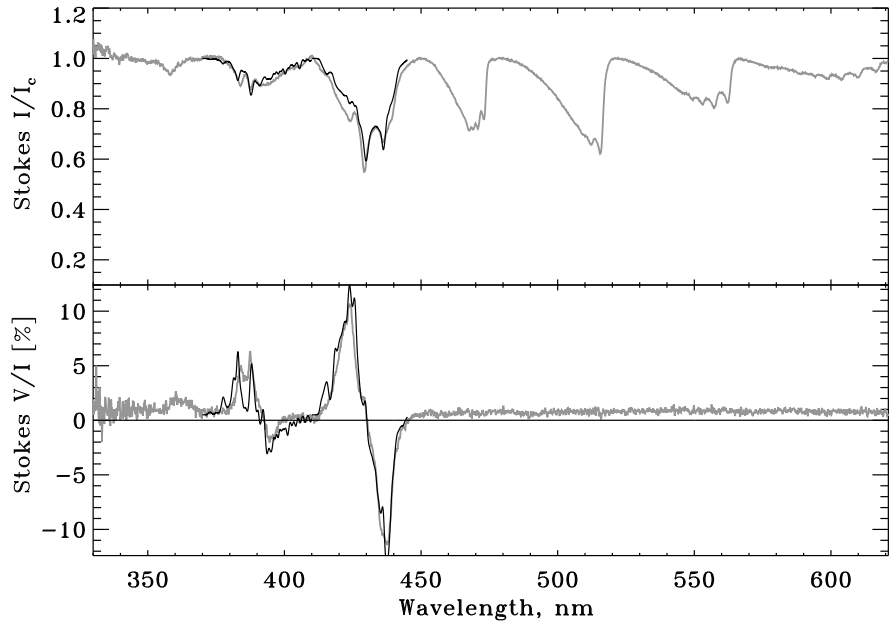}}
\centering
\caption{Intensity (top) and Stokes $V$ (bottom) of G99-37 observed in 2008. The notation is the same as in Figure~\ref{fig:GJ841B}.} 
\label{fig:G9937}
\end{figure}

The new observations of G99-37 were obtained with a higher S/N. As we mentioned the spectra from 2005 and 2008 were nearly identical, so we analyzed only the later one. The observed intensity and Stokes $V$ spectra are shown in Figure~\ref{fig:G9937} together with our model fit. The prominent polarized features are the two CH systems. The antisymmetric Stokes $V$ profile of the $A$--$X$ system ranges from about +10\% to $-$11\%. In the $B$--$X$ system, polarization reaches 5\%--6\%. The continuum polarization is about +1\%. There is also a new polarized feature of about +1\% above the continuum at 360\,nm, which was barely seen in the 2003 spectrum due to a larger noise level. It corresponds to the absorption band at the same wavelength. As modeled by \citet{ber07} the polarization cannot be due to the C$_2$ Deslandres-d'Azambuja bands $\Delta v =1$. We suppose that it may be caused by the CH $C^2\Sigma^+$--$X^2\Pi$ system. The latter is similar to the $B$--$X$ system in terms of the electronic quantum numbers and molecular constants, so their polarization signatures can be expected to be  similar too. Indeed, the overall sign of polarization is positive for both, while the weakness  of the $C$--$X$ system can be  explained by a higher excitation energy. A detailed modeling is required to confirm this interpretation. This task is however beyond the scope of the present Letter and will be carried out in a separate paper.

\begin{figure}
\resizebox{12.0cm}{!}{\includegraphics{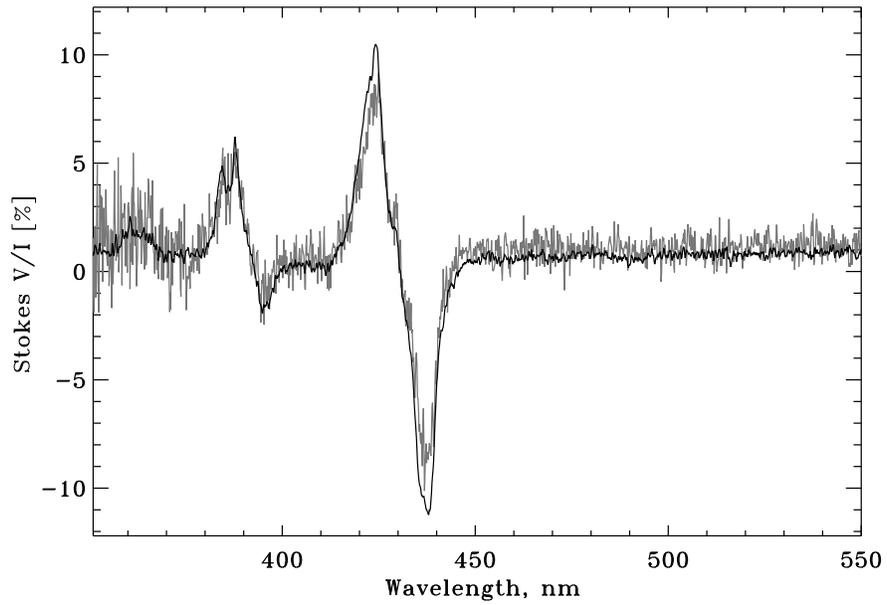}}
\centering
\caption{Circular polarization spectra of G99-37 from 2003 \citep{ber07} (in gray) and from 2008 (in black).}
\label{fig:G9937comp}
\end{figure}

As before, we searched for the best-fit solution simultaneously for Stokes $I$ and $V$ of the two CH systems. We found for the longitudinal field $B=7.3\pm0.3$\,MG and temperature $T=6200\pm200$\,K. These values are in a very good agreement with the results obtained by \citet{ber07}, but small discrepancies with the observations are seen at this better S/N. To get a better agreement between the model and observations we need perhaps to include more C$_2$ lines and abandon the assumption on a homogeneous field. In particular, a dipole model was apparently better for the $B$--$X$ system \citep{ber07}.

No change in the spectral features between the observations of 2005 and 2008 suggests that G99-37 is either a very fast or a very slow rotator. There are also no significant variations at the noise level of 1\%\ in individual polarization spectra recorded with the time interval of about 5 minutes. Hence, either the rotation period would have to be shorter than that or variations are smaller than 1\%. Alternatively, as there is no change in polarization between the 2005 and 2008 observations, the rotation period could be longer than three years. 
In fact, the circular polarization spectrum
appears to be almost unchanged from the 1973 measurement reported by
\citet{ang74}, when the difference in resolving power (about 30 in
the 1973 data) is taken into account. This suggests that the period of
rotation may be very long indeed, more than 100 years (or that the field
has an axis of symmetry fairly aligned with the rotation axis).
Indeed, it was shown by \citet{ber07} that the magnetic dipole on this WD points towards the Earth. If this coincides with the direction of the rotational axis, we will not be able to detect any rotational modulation on any time scale, so the period will remain unknown.

To compare the circular polarization spectra from 2008 to those obtained with ALFOSC at the Nordic Optical Telescope in 2003 \citep{ber07}, we plot them in Figure~\ref{fig:G9937comp}. Even though the spectral resolution of the 2008 spectrum ($R$=780) is somewhat higher than that of the 2003 spectrum ($R$=330) and the data of 2003 are noisier than our new data, it seems that there is a difference in the peaks of the $A$--$X$ system, which appear stronger by 1\%--2\%\ in 2008. However, we emphasize that these data were taken with different instruments and we see no variations in the 2005 and 2008 data obtained with the same instrument. Hence, the difference between the 2003 and 2005/2008 data may be due to an instrumental effect which was unaccounted in the calibration procedures. If this difference is real, the fact that the profile shapes and positions of the peaks have not altered indicates that either the abundance of the CH molecules has slightly increased in 2008 or orientation of the magnetic field changed, but the magnetic field strength stayed the same. Perhaps a higher spectral resolution is needed to reveal possible subtle variations in the spectra of G99-37.

\section{Conclusions}

We have presented observations of the cool magnetic DQ WD GJ 841B which is the second one showing polarization in the CH molecular bands and the first one unambiguously showing polarization in the C$_2$ Swan bands \citep[the case of peculiar polarized DQ WDs like LP790-29 and LHS2229 is still open for debate, see][]{hal08}. From our model fit we have found a magnetic field strength of 1.3 MG and a temperature of 6100K on the surface of GJ 841B. \citet{sch95} made a detailed equilibrium analysis of H/He/C mixtures under the physical conditions in the atmospheres of several peculiar DQ WDs and argued in favor of the presence of C$_2$H molecule in their photospheres. According to \citet{sch95}, the apparent absence of CH lines in their spectra is due to the low temperature ($\leq$ 7000K), high atmospheric pressure (log $P$ $\geq$ 9.0), and high carbon abundance (10$^{-3}$ or higher). It is worth to mention that GJ 841B has strong C$_2$ lines, low temperature, and yet shows the presence of CH without any signs of spectral features which can be attributed to C$_2$H. We also emphasize that CH in GJ 841B was discovered thanks to high S/N spectropolarimetry and its magnetism: the polarization signature of the $A$--$X$ band seen at 430 nm looks very similar to the strongly polarized CH absorption in the well-known G99-37. Some other non-magnetic DQ WDs may also have non-detected spectroscopic signatures of CH bands in the blue spectral region blended with C$_2$ lines. A possible lack of magnetism in those objects and difficulties with the adequate modeling of C$_2$ bands, however, make this detection rather difficult.   

We have also re-observed G99-37, historically the first discovered WD with polarized CH molecular bands. We have determined a magnetic field strength of 7.3 MG and a temperature of 6200 K for it, in a very good agreement with the previous result by \citet{ber07}. A comparison between the measurements in 2003, 2005, and 2008 reveals no unambiguous variations in the spectra.

To find more objects of this kind, we continue our survey of cool DQ WDs. We have also observed several DQ WDs that do not show polarization at a detectable level. These results will be presented in a forthcoming paper. They will help us set limits of a magnetic field in this type of objects.

\acknowledgments
The authors thank the referee for valuable input that helped to improve the paper greatly and especially for the discussion on the early work done on G99-37 in the 70's. This paper is based on observations made with the Nordic Optical Telescope and with VLT/FORS1. The authors take this chance to thank the wonderful staff of FORS1 at ESO whose work made so many wonderful discoveries possible. The Nordic Optical Telescope is operated on the island of La Palma jointly by Denmark, Finland, Iceland, Norway, and Sweden, in the Spanish Observatorio del Roque de los Muchachos of the Instituto de Astrofisica de Canarias. This work is supported by the Academy of Finland, grant 115417. SVB acknowledges the EURYI Award from the ESF (see www.esf.org/euryi) and the SNF grant PE002-104552.



{\it Facilities:} \facility{VLT:Kueyen (FORS1)}

\clearpage

\end{document}